\documentclass[12pt]{article}
\usepackage{amsmath}  % Better maths support
\begin{document}
%\input feynman

%\special{papersize=8.26in,11.69in}
%\textwidth15.0cm
%\textheight22.0cm
%\baselineskip1.0cm
%\setlength{\topmargin}{-1cm}
%\addtolength{\textheight}{1cm}
%\oddsidemargin+1.2cm
%\evensidemargin-1.2cm

%%%%%%%%%%%%%%% insert actual file mydefs.sty %%%%%%%%%%%%%%%%%%%%%%
%%%%%%%%%%%%%%%%%%%%%%%%%%%%%%%%%%%%%%%%%%%%%%%%%%%%%%%%%%%%%%%%%%%%%%%
%------------------------------------------------------------------------     
% MATH SYMBOLS
%
%fractions
\def\half{{1\over 2}}
\def\third{{1\over3}}
\def\fourth{{1\over4}}
\def\fifth{{1\over5}}
\def\sixth{{1\over6}}
\def\seventh{{1\over7}}
\def\eigth{{1\over8}}
\def\ninth{{1\over9}}
\def\tenth{{1\over10}}
\def\bN{\mathop{\bf N}}
\def\R{{\rm I\!R}}
\def\Eins{{\mathchoice {\rm 1\mskip-4mu l} {\rm 1\mskip-4mu l}
{\rm 1\mskip-4.5mu l} {\rm 1\mskip-5mu l}}}
\def\Z{{\mathchoice {\hbox{$\sf\textstyle Z\kern-0.4em Z$}}
{\hbox{$\sf\textstyle Z\kern-0.4em Z$}}
{\hbox{$\sf\scriptstyle Z\kern-0.3em Z$}}
{\hbox{$\sf\scriptscriptstyle Z\kern-0.2em Z$}}}}
\def\abs#1{\left| #1\right|}
\def\com#1#2{
        \left[#1, #2\right]}
\def\square{\kern1pt\vbox{\hrule height 1.2pt\hbox{\vrule width 1.2pt
   \hskip 3pt\vbox{\vskip 6pt}\hskip 3pt\vrule width 0.6pt}
   \hrule height 0.6pt}\kern1pt}
% \contract is a differential geometry contraction sign _|
\def\contract{\makebox[1.2em][c]{
        \mbox{\rule{.6em}{.01truein}\rule{.01truein}{.6em}}}}
\def\ltap{\ \raisebox{-.4ex}{\rlap{$\sim$}} \raisebox{.4ex}{$<$}\ }
\def\gtap{\ \raisebox{-.4ex}{\rlap{$\sim$}} \raisebox{.4ex}{$>$}\ }
\def\mn{{\mu\nu}}
\def\rs{{\rho\sigma}}
\newcommand{\Det}{{\rm Det}}
\def\Tr{{\rm Tr}\,}
\def\tr{{\rm tr}\,}
\def\sumij{\sum_{i<j}}
\def\e{\,{\rm e}}
%derivatives
\def\pa{\partial}
\def\dA{\partial^2}
\def\ddx{{d\over dx}}
\def\ddt{{d\over dt}}
\def\der#1#2{{d #1\over d#2}}
\def\lie{\hbox{\it \$}} % fancy L for the Lie derivative
\def\partder#1#2{{\partial #1\over\partial #2}}
\def\secder#1#2#3{{\partial^2 #1\over\partial #2 \partial #3}}
%
%equations
\newcommand{\be}{\begin{equation}}
\newcommand{\ee}{\end{equation}\noindent}
\newcommand{\bear}{\begin{eqnarray}}
\newcommand{\ear}{\end{eqnarray}\noindent}
\newcommand{\benn}{\begin{enumerate}}
\newcommand{\enn}{\end{enumerate}}
\newcommand{\veject}{\vfill\eject}
%
%reference to equations
\def\eq#1{{eq. (\ref{#1})}}
\def\eqs#1#2{{eqs. (\ref{#1}) -- (\ref{#2})}}
%
%integrals
\def\totint{\int_{-\infty}^{\infty}}
\def\posint{\int_0^{\infty}}
\def\negint{\int_{-\infty}^0}
\def\pint{{\dps\int}{dp_i\over {(2\pi)}^d}}
%
% PHYS SYMBOLS
\newcommand{\GeV}{\mbox{GeV}}
\def\FFdual{F\cdot\tilde F}
\def\bra#1{\langle #1 |}
\def\ket#1{| #1 \rangle}
\def\braket#1#2{\langle {#1} \mid {#2} \rangle}
\def\vev#1{\langle #1 \rangle}
\def\rightvac{\mid 0\rangle}
\def\leftvac{\langle 0\mid}
\def\ihbar{{i\over\hbar}}
% dirac matrix stuff
\def\ge{\hbox{$\gamma_1$}}
\def\gz{\hbox{$\gamma_2$}}
\def\gd{\hbox{$\gamma_3$}}
\def\go{\hbox{$\gamma_1$}}
\def\gt{\hbox{\$\gamma_2$}}
\def\gth{\hbox{$\gamma_3$}} 
\def\gf{\hbox{$\gamma_5\;$}}
\def\slash#1{#1\!\!\!\raise.15ex\hbox {/}}
\newcommand{\slD}{\raise.15ex\hbox{$/$}\kern-.57em\hbox{$D$}}
\newcommand{\slpartial}{\raise.15ex\hbox{$/$}\kern-.57em\hbox{$\partial$}}
\newcommand{\cL}{\cal L}
\newcommand{\D}{\cal D}
\newcommand{\Dhalf}{{D\over 2}}
\def\eps{\epsilon}
\def\epshalf{{\epsilon\over 2}}
\def\lag{( -\partial^2 + V)}
%worldline
\def\freeexp{{\rm e}^{-\int_0^Td\tau {1\over 4}\dot x^2}}
\def\kinb{{1\over 4}\dot x^2}
\def\kinf{{1\over 2}\psi\dot\psi}
\def\expk{{\rm exp}\biggl[\,\sum_{i<j=1}^4 G_{Bij}k_i\cdot k_j\biggr]}
\def\expp{{\rm exp}\biggl[\,\sum_{i<j=1}^4 G_{Bij}p_i\cdot p_j\biggr]}
\def\expshort{{\e}^{\half G_{Bij}k_i\cdot k_j}}
\def\expabb{{\e}^{(\cdot )}}
\def\epseps#1#2{\varepsilon_{#1}\cdot \varepsilon_{#2}}
\def\epsk#1#2{\varepsilon_{#1}\cdot k_{#2}}
\def\kk#1#2{k_{#1}\cdot k_{#2}}
\def\G#1#2{G_{B#1#2}}
\def\Gp#1#2{{\dot G_{B#1#2}}}
\def\GF#1#2{G_{F#1#2}}
\def\Dab{{(x_a-x_b)}}
\def\Dsq{{({(x_a-x_b)}^2)}}
\def\PITD{{(4\pi T)}^{-{D\over 2}}}
\def\4piTD{{(4\pi T)}^{-{D\over 2}}}
\def\4piT4{{(4\pi T)}^{-2}}
\def\TintmD{{\dps\int_{0}^{\infty}}{dT\over T}\,e^{-m^2T}
    {(4\pi T)}^{-{D\over 2}}}
\def\Tintm4{{\dps\int_{0}^{\infty}}{dT\over T}\,e^{-m^2T}
    {(4\pi T)}^{-2}}
\def\Tintm{{\dps\int_{0}^{\infty}}{dT\over T}\,e^{-m^2T}}
\def\Tint{{\dps\int_{0}^{\infty}}{dT\over T}}
\def\np{n_{+}}
\def\nm{n_{-}}
\def\Np{N_{+}}
\def\Nm{N_{-}}
\newcommand{\slG}{{{\dot G}\!\!\!\! \raise.15ex\hbox {/}}}
\newcommand{\Gd}{{\dot G}}
\newcommand{\Gund}{{\underline{\dot G}}}
\newcommand{\Gdd}{{\ddot G}}
\def\GBd12{{\dot G}_{B12}}
\def\Dx{\dps\int{\cal D}x}
\def\Dy{\dps\int{\cal D}y}
\def\Dpsi{\dps\int{\cal D}\psi}
\def\dint#1{\int\!\!\!\!\!\int\limits_{\!\!#1}}
\def\ddtau{{d\over d\tau}}
\def\ie{\hbox{$\textstyle{\int_1}$}}
\def\iz{\hbox{$\textstyle{\int_2}$}}
\def\id{\hbox{$\textstyle{\int_3}$}}
\def\ldop{\hbox{$\lbrace\mskip -4.5mu\mid$}}
\def\rdop{\hbox{$\mid\mskip -4.3mu\rbrace$}}
%
%VARIOUS
\newcommand{\1}{{\'\i}}
\newcommand{\no}{\noindent}
\def\non{\nonumber}
\def\dps{\displaystyle}
\def\sy{\scriptscriptstyle}
\def\sy{\scriptscriptstyle}
%-------------------------------------------------------------------------
%----------------------------------------------------------
% Title page
%\begin{document}
\pagestyle{empty}
\renewcommand{\thefootnote}{\fnsymbol{footnote}}
%\hskip 9cm {\sl LAPTH-}
\vskip-.1pt
%\hskip 9cm {\sl UMSNH-Phys/01-}   
%\vskip-.1pt
%\hskip 10cm hep-th/ 
\vskip .4cm
\begin{center}
{\Large\bf Closed-form weak-field expansion of \\
two-loop Euler-Heisenberg Lagrangians}
\vskip1.3cm
 {\large Gerald V. Dunne and Adolfo Huet} 
\\[1.5ex]
{\it
Department of Physics,
University of Connecticut, Storrs, CT 06269, USA
}
%{\it
%CSSM, Department of Physics, University of Adelaide,
%SA 5005, Australia\\
%and\\
%Department of Physics,
%University of Connecticut, Storrs, CT 06269, USA
%}
%\vskip .5cm
% {\large  Adolfo Huet} 
%\\[1.5ex]
%{\it
%Department of Physics,
%University of Connecticut, Storrs, CT 06269, USA
%}

\vskip.5cm
{\large David Rivera}
\\[1.5ex]
{\it
Department of Physics and Geology,
University of Texas Pan American, Edinburg, TX 78541-2999, USA}
\vskip.5cm

{\large Christian Schubert}
\\[1.5ex]
{\it
Instituto de F\'{\i}sica y Matem\'aticas
\\
Universidad Michoacana de San Nicol\'as de Hidalgo\\
Apdo. Postal 2-82\\
C.P. 58040, Morelia, Michoac\'an, M\'exico\\
schubert@ifm.umich.mx\\
}
\vspace{1.5cm}

{\large \bf Abstract}
\end{center}
\begin{quotation}
\noindent
We obtain closed-form expressions, in terms of the Faulhaber numbers,
for the weak-field expansion coefficients of the two-loop Euler-Heisenberg
effective Lagrangians in a magnetic or electric field. This follows from the
observation that the magnetic worldline Green's function has a natural expansion
in terms of the Faulhaber numbers.
\end{quotation}
\clearpage
\renewcommand{\thefootnote}{\protect\arabic{footnote}}
\pagestyle{plain}
%------------------------------------------------------

\setcounter{page}{1}
\setcounter{footnote}{0}

\section{Introduction: The Euler-Heisenberg Lagrangian at one and two loops}
\renewcommand{\theequation}{1.\arabic{equation}}
\setcounter{equation}{0}

The Euler-Heisenberg Lagrangian \cite{eulhei,weisskopf}
describes the effect of a virtual electron - positron pair
on an external Maxwell field in the one loop and constant field
approximation. Its standard proper time representation is
\bear
{\cal L}^{(1)}_{\rm spin}(F)&=& - {1\over 8\pi^2}
\int_0^{\infty}{dT\over T^3}
\,\e^{-m^2T} 
\biggl[
{(eaT)(ebT)\over \tanh(eaT)\tan(ebT)} 
\nonumber\\&&\hspace{70pt}
- {1\over 3}(a^2-b^2)T^2 -1
\biggr].
\label{eh1spin}
\ear
Here $T$ is the proper-time of the loop fermion, $m$ its mass, and $a,b$ are the two
Maxwell invariants, related to $\bf E$, $\bf B$ by $a^2-b^2 = B^2-E^2,\quad ab = {\bf E}\cdot {\bf B}$.
The superscript "(1)" stands for one loop.
A similar representation exists for scalar QED \cite{weisskopf,schwinger51}:
\bear
{\cal L}^{(1)}_{\rm scal}(F)&=&  {1\over 16\pi^2}
\int_0^{\infty}{dT\over T^3}
\,\e^{-m^2T} 
\biggl[
{(eaT)(ebT)\over \sinh(eaT)\sin(ebT)} 
\nonumber\\&&\hspace{60pt}
+{1\over 6}(a^2-b^2)T^2 -1
\biggr].
\label{eh1scal}
\ear
The Lagrangians (\ref{eh1spin}), (\ref{eh1scal})
historically provided the first examples for the concept of an
effective Lagrangian, and moreover the first nonperturbative
result in quantum field theory. See \cite{geraldrev} for a review of their
many applications and generalizations.

The proper time integrals in these formulas can be done exactly
in terms of certain special functions \cite{geraldrev}. Alternatively,
one can expand the integrands as power series in the field invariants,
using the Taylor expansions
\bear
{z \over\ \tanh(z)} &=& \sum_{n=0}^{\infty}{{\cal B}_{2n}\over
(2n)!} \,(2z)^{2n}\label{taylcoth2},\\
{z\over \sinh(z)} &=&
-\sum_{n=0}^{\infty}\Bigl(1-2^{1-2n}\Bigr){{\cal B}_{2n}\over (2n)!}\,(2z)^{2n}.
\label{taylcsch}
\ear
Here the ${\cal B}_{2n}$ are the Bernoulli numbers.
The terms in this expansion involving $N=2n$ powers of the field
contain the information on the low energy limits of the $N$ photon scattering
amplitudes. Thus in the low energy limit 
one can obtain these amplitudes in closed form \cite{mascvi}. 
%This should be
%contrasted with the fact that the full $N$ photon amplitudes are presently known
%only for the four point case \cite{}, and even this only on-shell or with maximally
%two legs off-shell \cite{}. 

%A more subtle application of the same power series expansion uses Borel analysis
%\cite{dunhal,dunsch1}. 
\noindent
For a purely magnetic field 
eqs. (\ref{eh1spin}), (\ref{eh1scal}), (\ref{taylcoth2}), (\ref{taylcsch}) yield
\begin{eqnarray}
{\cal L}^{(1)}_{\rm spin}(B)\!\!\!&=&\!\! -8\left(\frac{\alpha}{\pi}\right) B^2 
\sum_{n=0}^\infty {2^{2n} {\cal B}_{2n+4}\over (2n+4)(2n+3)(2n+2)}\left(
\frac{eB}{m^2}\right)^{2n+2} ,
\label{1lwspin}\\
{\cal L}^{(1)}_{\rm scal}(B)\!\!\! &=& \! 4\left(\frac{\alpha}{\pi}\right) B^2
\sum_{n=0}^\infty { 2^{2n}(2^{-2n-3}-1) {\cal B}_{2n+4}\over
(2n+4)(2n+3)(2n+2)}\left(
\frac{eB}{m^2}\right)^{2n+2} .
\label{1lwscal}
\end{eqnarray}
%\begin{eqnarray}
%{\cal L}^{(1)}_{\rm spin}(B)&=&-\frac{2m^4}{\pi^2}\left(\frac{eB}{m^2}\right)^4
%\sum_{n=0}^\infty \frac{2^{2n} {\cal B}_{2n+4}}{(2n+4)(2n+3)(2n+2)}\left(
%\frac{eB}{m^2}\right)^{2n}
%\label{1lwspin}\\
%{\cal L}^{(1)}_{\rm scal}(B)&=&\frac{m^4}{\pi^2}\left(\frac{eB}{m^2}\right)^4
%\sum_{n=0}^\infty \frac{ 2^{2n}(2^{-2n-3}-1) {\cal B}_{2n+4}}{
%(2n+4)(2n+3)(2n+2)}\left(
%\frac{eB}{m^2}\right)^{2n}
%\label{1lwscal}
%\end{eqnarray}
An analysis of the coefficients shows
that these series are divergent but Borel summable \cite{chadha,dunhal,dunsch1}.
Moreover, the summability relates to the fact that the
effective Lagrangian is real in the magnetic case. The corresponding series for the
purely electric case is obtained by the replacement
$B^2 \to -E^2$, which turns the alternating
series into a non-alternating one. The non-alternating
series is not Borel summable,  
which is an indication of the well-known fact
that the electric effective Lagrangian has an
imaginary part. 
Although this imaginary part is nonperturbative in nature,
it is possible to calculate it from the expansion coefficients 
by an analysis of their large $n$ behaviour, combined with
a Borel dispersion relation \cite{chadha,dunhal,dunsch1}.

This paper addresses the possibility of analyzing such weak-field expansions at higher loop
orders in QED. The two loop corrections to the Lagrangians (\ref{eh1spin}),
(\ref{eh1scal}) have first been considered by Ritus, who 
obtained them in terms of integral representations 
both in spinor \cite{ritspin,ginzburg} and scalar QED \cite{ritscal}. 
Other representations for the same Lagrangians were later given in
\cite{ditreu,rescsc,frss,report}, however they all involve the same apparently
intractable type of double integrals. As a consequence, at the two loop level
presently only the first few coefficients of the power series expansions
for ${\cal L}^{(2)}_{\rm spin/scal}(B)$
are known \cite{dunsch1}, and the imaginary parts
${\rm Im}{\cal L}^{(2)}_{\rm spin/scal}(E)$
are known explicitly only in the leading weak field limit \cite{lebrit,dunsch1}
(although Lebedev and Ritus succeeded in establishing the general structure
of the imaginary part \cite{lebrit}).

This should be contrasted with the case of a self-dual (euclidean) field,
corresponding to external photon lines of definite helicity,
where the two-loop
effective Lagrangians can be obtained in closed form \cite{sd,sd1,sd2}.
This result allows one to extend the 
aforementioned calculation of the low energy limit of the
$N$ photon amplitudes to the two loop level for one particular
component of the photon S matrix, the so-called MHV amplitudes
\cite{sd1}.

For general fields,
it seems difficult to make further progress at the two loop level
without having a closed form expression for the weak field
expansion coefficients. It is the purpose of the present paper
to derive such expressions for the purely magnetic (or purely electric)
cases.
Thus, our goal is to obtain the coefficients of the weak field expansion of the (renormalized)
magnetic two-loop effective Euler-Heisenberg Lagrangians 
${\cal L}^{(2)}_{\rm spin/scal}(B)$,
\bear
{\cal L}^{(2)}_{\rm spin}(B) &=&
\left(\frac{\alpha}{4\pi}\right)^2 B^2
\sum_{n=0}^{\infty}a_{\rm spin}^{(2)}(n)\Bigl(\frac{eB}{m^2}\Bigr)^{2n+2},
\label{L2spinexp}\\
{\cal L}^{(2)}_{\rm scal}(B) &=&
-\frac{1}{2}\left(\frac{\alpha}{4\pi}\right)^2 B^2
\sum_{n=0}^{\infty}a_{\rm scal}^{(2)}(n)\Bigl(\frac{eB}{m^2}\Bigr)^{2n+2}.
\label{L2scalexp}
\ear
Note the difference in normalization between the scalar and spinor QED cases,
which takes into account 
the global factor of $-2$ from statistics and degrees of freedom.

\section{Two loop expansion coefficients for spinor QED}
\renewcommand{\theequation}{2.\arabic{equation}}
\setcounter{equation}{0}
We start with the following integral representation of the on-shell
renormalized Lagrangian, which was obtained in 
\cite{frss}:
\bear
{\cal L}_{\rm spin}^{(2)}(B)
&=& 
{\cal L}_{\rm spin,main}^{(2)}(B)
+{\cal L}_{\rm spin,{\delta m}}^{(2)}(B) .
\label{splitLspin}
\ear
Here the ``main part'' is given  by the following two parameter integral,
\bear
{\cal L}_{\rm spin,main}^{(2)}(B)=
\left(\frac{\alpha}{4\pi }\right)^2 B^2
\int_{0}^{\infty}\frac{dz}{z^3}e^{-\frac{m^2}{eB}\, z}
\int_0^1 du
\,
\Bigl[
L(z,u)-L_{02}(z,u)-\frac{g(z)}{G}
\Bigr] \non\\ \label{L2spinmain}
\ear
where $z=eBT$ and
\bear
L(z,u)&=&\frac{z}{\tanh(z)}
\Biggl\lbrace
B_1
{\ln ({G/G^z})
\over{(G-G^z)}^2}
+{B_2\over
G^z(G-G^z)}
+{B_3\over
G(G-G^z)}
\Biggr\rbrace
\, ,
\nonumber\\
B_1&=&4z
\Bigl(
\coth(z)-\tanh(z)
\Bigr)
G^z-4G
\, ,
\quad  \nonumber\\
B_2&=&2\dot G\dot G^z+ 
z(8\tanh(z)-4\coth(z))
G^z-2
\, ,
\quad \nonumber\\
B_3&=&4G
-2\dot G\dot G^z
-4z\tanh(z)G^z+2\, ,\nonumber\\
L_{02}(z,u)&=&-{12\over G}+2z^2\, ,\nonumber\\
g(z)&=&-6\biggl[
{z^2\over{\sinh(z)}^2}+z\coth(z)-2
\biggr]\,\, .
\label{L2spinmainint}
\ear\no
The integrand involves the 
so-called worldline Green's function $G(u)$ and the ``magnetic''
Green's function $G^z(z,u)$, as well as their $u$-derivatives (denoted as $\dot{G}$
and $\dot{G}^z$, respectively):
\bear
G &=& u(1-u) \, ,\non\\
\dot G &=& 1-2u \, , \non\\
G^z &=& \half {\cosh (z)-\cosh(z\dot G)\over
z\sinh (z)} \, ,
\non\\
\dot G^z &=& {\sinh(z\dot G)\over\sinh(z)}
\,\, .
\label{defG}
\ear
The function $L(z,u)$ is essentially the two-loop integrand before renormalization.
The term $L_{02}(z,u)$ removes the
order $z^0, z^2$ terms, 
which implements the renormalization of
the charge and the field, and the removal of
the vacuum energy.
(In the following we will not always make this subtraction explicit.)
The other subtraction term involving $g(z)\over G$ relates to mass renormalization,
which is necessary starting at the two loop level.
In the worldline formalism the need for this mass renormalization subtraction
can be recognized from the appearance of singularities at $u=0$ and $u=1$ 
\cite{rescsc,frss}. Those singular  
terms can be absorbed into a mass shift of the one-loop lagrangian,
$\delta m_0{\partial\over \partial m_0}
{\cal L}_{\rm spin}^{(1)}(B)$,
however this leaves a finite remainder 
${\cal L}_{\rm spin, {\rm \delta m}}^{(2)}(B)$,
\bear
{\cal L}_{\rm spin, {\rm \delta m}}^{(2)}(B)
&=&
-{\alpha\over {(4\pi)}^3}
e B m^2 \int_0^{\infty}
{dz\over z^2}\,{\rm e}^{-\frac{m^2}{eB}\, z}
\biggl[
{z\over\tanh(z)}
-{z^2\over 3}-1
\biggr]
\non\\&&\qquad
\times
\biggl[18-12\gamma-
12\ln \left(\frac{m^2 z}{eB}\right)+12\frac{eB}{m^2 z}
\biggr].
\label{L2spindm}
\ear
The weak field expansion coefficients of 
${\cal L}_{\rm spin,{\rm \delta m}}^{(2)}(B)$ 
are again easy to obtain using the
Taylor expansion (\ref{taylcoth2}). One finds
\bear
{\cal L}^{(2)}_{\rm spin,{\rm \delta m}}(B) &=&
\left(\frac{\alpha}{4\pi}\right)^2 B^2
\sum_{n=0}^{\infty}a_{\rm spin,{\rm \delta m}}^{(2)}(n)
\Bigl({eB\over m^2}\Bigr)^{2n+2},\non\\
a_{\rm spin,{\rm \delta m}}^{(2)}(n)
&=& -12 {2^{2n+4}{\cal B}_{2n+4}\over (2n+4)(2n+3)}
\Bigl({3\over 2}-\gamma -\psi(2n+2)\Bigr).
\label{cndm}
\ear

However, the main integral term (\ref{L2spinmain}) is much more difficult to expand.
A brute-force expansion was done in \cite{dunsch1}, but no closed-form expression was
obtained. The technical challenge is to expand the integrand (excluding the
$e^{-\frac{m^2}{eB}\, z}$ factor) in a series in $z$, in such a way that the $u$
integrals can also be done easily. This is complicated by the fact that the magnetic
Green's function $G^z(z,u)$ couples these two parametric variables in a nontrivial
manner. The main observation of this paper is that there exists an expansion of
$G^z(z, u)$ which decouples the variables in an elegant manner. Remarkably, $G^z(z,
u)$ is essentially the generating function of the so-called Faulhaber numbers of
number theory \cite{gesvie}. Namely,
\bear
G^z(z, u) &=& \half {\cosh (z)-\cosh(z\dot G)\over
z\sinh (z)}\non\\
&=&
G(u)-\sum_{m=1}^{\infty}(2z)^{2m}\sum_{k=1}^m
\tilde f(m,k)\, [G(u)]^{k+1} \quad .
\label{expfaulhaber}
\ear
Here we have defined 
\bear
\tilde f(m,k) \equiv {(-1)^{k+1}\over (2m+1)!}f(m,k)
\label{defftilde}
\ear
with $f(m,k)$ the Faulhaber numbers. These numbers
can in turn be written as linear combinations
of the Bernoulli numbers \cite{gesvie}:
\bear
f(m,k) = (-1)^{k+1}\sum_{j=0}^{\lfloor(k-1)/2\rfloor}
{1\over k-j}{2k-2j\choose k+1}{2m+1\choose 2j+1}{\cal B}_{2m-2j}
\label{faulhabertobernoulli}
\ear
($m,k\geq 1$).

Given such an expansion, the $u$ integrals are now trivial as they involve powers of
the free Green's function $G(u)=u(1-u)$. Using the standard Euler
beta function integral we define
\bear
\int_0^1 du\,\bigl[u(1-u)\bigr]^n &=& {n!^2\over (2n+1)!} \equiv \beta(n+1).
\label{eulerint}
\ear
To illustrate the Faulhaber expansion (\ref{expfaulhaber}), consider the first 
few terms:
\bear
G^z = G + \Bigl(1-z\coth(z)\Bigr)G^2 + O(G^3) =
G - \third G^2 z^2 + O(z^4,G^2)\, .
\label{expGzfirstfew}
\ear
This suggests 
rearranging $L(z,u)$ as a series in $\Delta G/G$ where $\Delta G = G^z - G$ is the
difference between the magnetic and free worldline Green's functions.
After simple manipulations, this leads to
\bear
L(z,u) &=&
\!\! \frac{4 z^2}{G}
+\sum_{i=0}^{\infty} \Bigl(-{\Delta G\over G}\Bigr)^i \biggl\lbrace
-{z^2 \over \sinh(z)^2} \frac{4}{G} {1\over (i+1)(i+2)}\non \\
&&-  {z\over \tanh(z)}{1\over G} \left( 8 + {4 \over i+2 } \right) +
{z\over \tanh(z)}  \frac{2 \, \dot G (\dot G_z - \dot G)}{G^2}
 \biggr\rbrace. \non\\
\label{rearranged} \ear
The $u$ derivatives appearing in the last term can be removed
by an integration by parts, yielding
\bear
L(z,u) &=&
\frac{4 z^2}{G}+ 4 {z\over\tanh(z)}\Bigl( {z\over\tanh(z)}-1\Bigr)\non\\
\hskip -2cm && \hskip -2cm +\sum_{i=0}^{\infty} \Bigl(-{\Delta G\over G}\Bigr)^i \biggl\lbrace
-{z^2 \over \sinh(z)^2} \frac{4}{G} {1\over (i+1)(i+2)}
 -{z\over \tanh(z)}{1\over G} \left( 8 + {4 \over i+2 } \right) 
 \biggr\rbrace\non\\&& \hspace{-40pt}
 -
{2z\over \tanh(z)}\sum_{i=1}^{\infty}
 \Bigl(-{\Delta G\over G}\Bigr)^{i} \, \frac{1}{i}
\, \biggl\lbrace \frac{2-4(i+1)}{G} + \frac{i+1}{G^2}\biggr\rbrace
\quad .
\label{Lpartint} 
\ear
Performing the subtraction, implied in (\ref{L2spinmain}), of all terms
$O({1\over G})$, we finally obtain
\bear
{\cal L}_{\rm spin,main}^{(2)}(B)
&=&
\left(\frac{\alpha}{4\pi}\right)^2 B^2
\int_0^{\infty}{dz\over z^3}
\,\e^{-\frac{m^2}{eB}\,z}
\int_0^1 du \,\,
l(z,G,G^z),
\label{rewriteL2spinmain}
\ear
where
\bear
l(z,G,G^z) &=&\sum_{i=1}^{\infty} \Bigl({-\Delta G\over G}\Bigr)^i \biggl\lbrace
-{z^2 \over \sinh(z)^2} \frac{4}{G} {1\over (i+1)(i+2)}
+{z\over \tanh(z)}{8\over G} {1\over i(i+2)} 
 \biggr\rbrace\non\\&&
\hskip -2cm - {2z\over \tanh(z)}
\sum_{i=2}^{\infty}
 \Bigl(-{\Delta G\over G}\Bigr)^{i} \, \left(1+{1\over i}\right){1\over G^2}
+{4\over G}{z\over \tanh(z)}
\Bigl( {\Delta G\over G^2}
+{z\over\tanh(z)}-1\Bigr)
\non\\
&& \hskip -2cm + 4 {z\over\tanh(z)}\Bigl({z\over\tanh(z)}-1\Bigr) \, .
\label{lspin}
\ear
Here (\ref{expfaulhaber}) was used to obtain the $1\over G$  subtraction for the
last term. We now expand the trigonometric functions using (\ref{taylcoth2}) and
\bear
{z^2 \over \sinh^2(z)} = \sum_{n=0}^{\infty}
{{\cal B}_{2n}\over (2n)!}(1-2n) (2z)^{2n} \quad .
\label{taylsinh2}
\ear
This leads directly to the following closed-form expression for the expansion
coefficients:
\bear
a_{\rm spin,main}^{(2)}(n)\!\!\! &=& 2^{2n+4}(2n+1)! 
\Biggl\lbrace
\sum_{i=1}^{n+2}\sum_{M=i}^{n+2}\sum_{{m_1,\ldots ,m_i = 1\atop
{\scriptstyle \sum m_i = M}}}^M \sum_{k_1=1}^{m_1}\tilde f(m_1,k_1)
\cdots \sum_{k_i=1}^{m_i} \tilde f(m_i,k_i)\,
\non\\&&
\hskip -2cm
\times
\beta\Bigl(\sum_{j=1}^ik_j\Bigr)
\biggl\lbrack
{4 \over (i+1)(i+2)}
\Bigl(2(n-M)+3\Bigr)
+{8\over i(i+2)}
\biggr\rbrack
t_{n+2-M}
\non \\
&& \hskip -3cm  -
\sum_{i=2}^{n+2}\sum_{M=i}^{n+2}\sum_{{m_1,\ldots ,m_i = 1\atop
{\scriptstyle \sum m_i = M}}}^M \sum_{k_1=1}^{m_1}\tilde f(m_1,k_1)
\cdots \sum_{k_i=1}^{m_i} \tilde f(m_i,k_i) \,\beta\Bigl(\sum_{j=1}^i
k_j - 1 \Bigr) {2  (i+1)\over i}
 \,t_{n+2-M}\non\\
&& \hskip -3cm -\, 4\sum_{m=2}^{n+2}\sum_{k=2}^m\tilde f(m,k)
 \beta(k-1)t_{n+2-m}
 -8 (n+2) t_{n+2}
 \Biggr\rbrace
 \label{big} \ear
where we have defined the short-hand:
\bear
t_n &\equiv&{{\cal B}_{2n}\over (2n)!}\,\,\, .
\label{defhn}
\ear
The full two-loop expansion coefficients in (\ref{L2spinexp}) are now given in
closed-form by combining (\ref{cndm}) and (\ref{big}):
\bear
a_{\rm spin}^{(2)}(n) &=& a_{\rm spin,main}^{(2)}(n) +a_{\rm spin,{\rm \delta m}}^{(2)}(n)
\,\, .
\label{cspintot}
\ear

\section{Expansion coefficients for scalar QED}
\renewcommand{\theequation}{3.\arabic{equation}}
\setcounter{equation}{0}

The case of scalar QED can be treated completely analogously, starting from
any of the various representations for ${\cal L}^{(2)}_{\rm scal}(B)$ given
in \cite{rescsc,frss,report}. 
We will give
here only the final result for the weak field 
expansion coefficients (as defined by (\ref{L2scalexp})):
\bear
a_{\rm scal}^{(2)}(n) &=& -2^{2n+4}(2n+1)!
\biggl\lbrace
\sum_{M=2}^{n+2}
\sum_{k=2}^{M}
\tilde f(M,k)\,\beta(k-1)\,s_{n+2-M}
\non\\&& 
\hskip -2cm 
+\sum_{i=2}^{n+2}\sum_{M=i}^{n+2}\sum_{{m_1,\ldots ,m_i = 1\atop 
{\scriptstyle \sum m_i = M}}}^M
\sum_{k_1=1}^{m_1}\tilde f(m_1,k_1)
\cdots \sum_{k_i=1}^{m_i}
\tilde f(m_i,k_i)\,\beta\Bigl(\sum_{j=1}^ik_j - 1\Bigr) \,s_{n+2-M}
\non\\&& \hskip -2cm
+\sum_{i=1}^{n+2}\sum_{M=i}^{n+2}\sum_{{m_1,\ldots ,m_i = 1\atop {\scriptstyle \sum m_i = M}}}^M
\sum_{k_1=1}^{m_1}\tilde f(m_1,k_1)
\cdots \sum_{k_i=1}^{m_i}
\tilde f(m_i,k_i)
\,\beta\Bigl(\sum_{j=1}^ik_j \Bigr)
\non\\&&\qquad\qquad\times
{4\over i+2}
\Bigl[{1\over i+1}-(2n-2M+3)\Bigr]
 \,s_{n+2-M}\non\\&&\hskip -2cm 
+\sum_{i=1}^{n+1}\sum_{M=i}^{n+1}\sum_{{m_1,\ldots ,m_i = 1\atop 
{\scriptstyle \sum m_i = M}}}^M
\sum_{k_1=1}^{m_1}\tilde f(m_1,k_1)
\cdots \sum_{k_i=1}^{m_i}
\tilde f(m_i,k_i)
\non\\&&\qquad\qquad\times
{2\over (i+1)(i+2)}
\,\beta\Bigl(\sum_{j=1}^ik_j +1\Bigr) \,s_{n-M+1}
\non\\&& \hskip -1cm
- s_{n+1}
+ 2 s_{n+2}\Bigl[2+(2n+2)\Bigl(4-3\psi(2n+3)-3\gamma\Bigr)\Bigr]
\biggr\rbrace.
\label{bigscal}
\ear
Here we have defined the short-hand notation:
\bear
s_n \equiv - \Bigl(1-2^{1-2n}\Bigr)
{{\cal B}_{2n}\over (2n)!}
\,\,\, .
\label{defsn}
\ear

\section{Conclusions}
\renewcommand{\theequation}{4.\arabic{equation}}
\setcounter{equation}{0}

We have found closed-form expressions for the weak field
expansion coefficients
of the two loop corrections to the renormalized Euler-Heisenberg Lagrangians
in a purely magnetic (or purely electric) field. 
As a check, we have verified that
eqs.(\ref{cspintot}) and (\ref{bigscal}) indeed reproduce the known 
low order coefficients in these expansions \cite{frss,dunsch1}:
\bear
{\cal L}_{\rm spin}^{(2)}[B]
&=&\left(\frac{\alpha}{4\pi}\right)^2
\frac{B^2}{81}
\Biggl[
64 
{\Bigl({eB\over m^2}\Bigr)}^2
-{1219\over 25}
{\Bigl({eB\over m^2}\Bigr)}^4
+ {135308\over 1225}
{\Bigl({eB\over m^2}\Bigr)}^6
-\ldots \,
%-{791384\over 1575}
%{\Bigl({eB\over m^2}\Bigr)}^{8}
%+\ldots \, 
\Biggr]
,
\non\\
{\cal L}_{\rm scal}^{(2)}[B]
&=&
\left(\frac{\alpha}{4\pi}\right)^2
\frac{B^2}{81}
\Biggl[
{\displaystyle \frac {275}{8}}
{\Bigl({eB\over m^2}\Bigr)}^2
-
{\displaystyle \frac {5159}{200}}
{\Bigl({eB\over m^2}\Bigr)}^4
+
{\displaystyle \frac {2255019}{39200}}
{\Bigl({eB\over m^2}\Bigr)}^6
-\ldots \,
%-
%{\displaystyle \frac {931061}{3600}}
%{\Bigl({eB\over m^2}\Bigr)}^{8}
%+\ldots \,
\Biggr]
.
\non\\
\label{lowexpand}
\ear
Although the closed-form formulae (\ref{cspintot}) and (\ref{bigscal}) are 
significantly
more complicated than the corresponding ones for the case of a self-dual
field \cite{sd1,sd2}, their structure is still similar insofar as
they can, using (\ref{faulhabertobernoulli}), be written in terms of 
folded sums of Bernoulli numbers with factorial coefficients. 
Of course, it is quite possible that these formulas can still be
simplified. 

We note that the two-loop expansion coefficients are still rational
numbers, after extracting a factor of $(\alpha/\pi)^2$, just as the one-loop
coefficients are rational after extracting a factor of $(\alpha/\pi)$.
A question of obvious interest is whether  
this property persists to higher loop orders. Based on a comparison with
what is known about the coefficients of the QED 
$\beta$ - functions \cite{rosner,jowiba,bender,gkls,brdekr}
we consider it likely that rationality will be found to hold 
at least for the quenched (order $N_f$) contributions to the 
Euler-Heisenberg Lagrangians at arbitrary
loop order.      

Finally, since in the worldline formalism the magnetic Green's function
$G^z$ is the basic ingredient appearing in the integral representations
for all processes involving constant magnetic fields
\cite{rescsc,report}, we expect the Faulhaber expansion (\ref{expfaulhaber})  
to become useful for other calculations of this type, including 
possibly higher loop orders.

\vskip15pt
{\bf Acknowledgements:}
G.~Dunne and C.~Schubert gratefully acknowledge the support of NSF and CONACyT
through a US-Mexico collaborative research grant, NSF-INT-0122615. D.~Rivera and
C.~Schubert thank the Louis Stokes Alliance for Minority Participation
for financial support. G.~Dunne and A.~Huet thank the DOE for support through grant 
DEFG02-92ER40716.

%\vfill\eject


\begin{thebibliography}{99}
\bibitem{eulhei}
W.~Heisenberg and H.~Euler, 
Z. Phys. {\bf 98}, 714 (1936); an English translation is available at physics/0605038. 
\bibitem{weisskopf}
V.~Weisskopf, K. Dan. Vidensk. Selsk. Mat. Fy. Medd. {\bf 14}, 1 (1936), reprinted in {\it
Quantum Electrodynamics}, J. Schwinger (ed.), Dover, New York 1958.
\bibitem{schwinger51}
J.~Schwinger, Phys. Rev. {\bf 82}, 664 (1951).
\bibitem{geraldrev}
G.~V.~Dunne, ``Heisenberg-Euler Effective Lagrangians: Basics and Extensions'', in
Ian Kogan Memorial Collection, {\it From Fields to Strings: Circumnavigating
Theoretical Physics}, Vol. I, M.A. Shifman et al. (Eds.), World Scientific, Singapore,
2004, 445 [hep-th/0406216].
\bibitem{mascvi}
L.~C.~Martin, C.~Schubert 
and V.~M.~Villanueva Sandoval, Nucl. Phys. {\bf B 668}, 335 (2003) [hep-th/0301022].
\bibitem{chadha}
S.~Chadha and P.~Olesen,
%``On Borel Singularities In Quantum Field Theory,''
Phys.\ Lett. {\bf B 72}, 87 (1977).
%%CITATION = PHLTA,B72,87;%%
\bibitem{dunhal}
G.~V.~Dunne and T.~M.~Hall, Phys. Rev. {\bf D 60} 065002 (1999) [hep-th/9902064].
\bibitem{dunsch1}
G.~V.~Dunne and C.~Schubert, Nucl. Phys. {\bf B 564} (2000) 591 [hep-th/9907190].
\bibitem{ritspin}
V.~I.~Ritus, 
Zh. Eksp. Teor. Fiz. {\bf 69} (1975) 1517 [Sov. Phys. JETP {\bf 42} (1975) 774].
\bibitem{ginzburg}
V.~I.~Ritus,
{\it The Lagrangian Function of
an Intense Electromagnetic Field and Quantum Electrodynamics
at Short Distances}, in {\it Proc. Lebedev Phys. Inst. vol. 168},
V. I. Ginzburg ed., Nova Science Publ., NY 1987. 
\bibitem{ritscal}
V.~I.~Ritus, Zh. Eksp. Teor. Fiz. {\bf 73} (1977) 807 [Sov. Phys. JETP {\bf 46} (1977) 423].
\bibitem{ditreu}
W.~Dittrich and M.~Reuter, {\it
Effective Lagrangians in Quantum Electrodynamics},
Springer 1985.
\bibitem{rescsc}
M.~Reuter, M.~G.~Schmidt and C.~Schubert, Ann. Phys. (N.Y.) {\bf 259} (1997) 313
[hep-th/9610191].
\bibitem{frss}
D.~Fliegner, M.~Reuter, M.~G.~Schmidt, C.~Schubert,
Theor. Math. Phys. {\bf 113} (1997) 1442 [hep-th/9704194].
\bibitem{report}
C.~Schubert, Phys. Rept. {\bf 355} (2001) 73 [hep-th/0101036].
\bibitem{lebrit}
S.~L.~Lebedev, V.~I.~Ritus, 
Zh. Eksp.
Teor. Fiz. {\bf 86} (1984) 408 [JETP {\bf 59} (1984) 237].
\bibitem{sd}
G.~V.~Dunne, C.~Schubert, Phys. Lett.  {\bf B 526} (2002) 55, [hep-th/0111134].
\bibitem{sd1}
G.~V.~Dunne, C.~Schubert, JHEP 0208:053 (2002) [hep-th/0205004].  
\bibitem{sd2}
G.~V.~Dunne, C.~Schubert, JHEP 0206:042 (2002) [hep-th/0205005]. 
\bibitem{gesvie}
I.~M.~Gessel and X.~G.~Viennot, "Determinants, Paths, and Plane Partitions", Preprint, 1989.
\bibitem{rosner}
J.~L.~Rosner, Phys. Rev. Lett. {\bf 17} (1966) 1190; Ann. Phys. {\bf 44} (1967) 11.
\bibitem{jowiba}
K.~Johnson, R.~Willey, and M.~Baker, Phys. Rev. {\bf 163} (1967) 1699.
\bibitem{bender}
C.~M.~Bender, R.~W.~Keener and R.~E.~Zippel,
%``New Approach To The Calculation Of F(1) (Alpha) In Massless Quantum
%Electrodynamics,''
Phys.\ Rev.  {\bf D 15}, 1572 (1977).
%%CITATION = PHRVA,D15,1572;%%
\bibitem{gkls}
S.~G.~Gorishny, A.~L.~Kataev, S.~A.~Larin, and L.~R.~Surguladze,
Phys. Lett. {\bf B 256} (1991) 81.
\bibitem{brdekr}
D.~J.~Broadhurst, R.~Delbourgo, and D.~Kreimer,
Phys. Lett. {\bf B 366} (1996) 421 [hep-ph/9509296].

\end{thebibliography}
\end{document}